\documentclass[aps,prd,twocolumn,10pt
,%preprint,
tightenlines,%superscriptaddress,
nofootinbib,showpacs]{revtex4-1}
\usepackage{amssymb,latexsym}
\usepackage{amsmath,amsbsy,bbm}
\usepackage{epsfig,bm}
\usepackage{graphicx,comment}
\DeclareMathOperator{\tr}{tr}

\begin{document}
\def\a{{\alpha}}
\def\b{{\beta}}
\def\d{{\delta}}
\def\D{{\Delta}}
\def\X{{\Xi}}
\def\e{{\varepsilon}}
\def\g{{\gamma}}
\def\G{{\Gamma}}
\def\k{{\kappa}}
\def\l{{\lambda}}
\def\L{{\Lambda}}
\def\m{{\mu}}
\def\n{{\nu}}
\def\o{{\omega}}
\def\O{{\Omega}}
\def\S{{\Sigma}}
\def\s{{\sigma}}
\def\th{{\theta}}

\def\ol#1{{\overline{#1}}}

\def\Dslash{D\hskip-0.65em /}
\def\Dtslash{\tilde{D} \hskip-0.65em /}

\def\CPT{{$\chi$PT}}
\def\QCPT{{Q$\chi$PT}}
\def\PQCPT{{PQ$\chi$PT}}
\def\tr{\text{tr}}
\def\str{\text{str}}
\def\diag{\text{diag}}
\def\order{{\mathcal O}}

\def\cF{{\mathcal F}}
\def\cS{{\mathcal S}}
\def\cC{{\mathcal C}}
\def\cB{{\mathcal B}}
\def\cT{{\mathcal T}}
\def\cQ{{\mathcal Q}}
\def\cL{{\mathcal L}}
\def\cO{{\mathcal O}}
\def\cA{{\mathcal A}}
\def\cQ{{\mathcal Q}}
\def\cR{{\mathcal R}}
\def\cH{{\mathcal H}}
\def\cW{{\mathcal W}}
\def\cM{{\mathcal M}}
\def\cD{{\mathcal D}}
\def\cN{{\mathcal N}}
\def\cP{{\mathcal P}}
\def\cK{{\mathcal K}}
\def\Qt{{\tilde{Q}}}
\def\Dt{{\tilde{D}}}
\def\St{{\tilde{\Sigma}}}
\def\cBt{{\tilde{\mathcal{B}}}}
\def\cDt{{\tilde{\mathcal{D}}}}
\def\cTt{{\tilde{\mathcal{T}}}}
\def\cMt{{\tilde{\mathcal{M}}}}
\def\At{{\tilde{A}}}
\def\cNt{{\tilde{\mathcal{N}}}}
\def\cOt{{\tilde{\mathcal{O}}}}
\def\cPt{{\tilde{\mathcal{P}}}}
\def\cI{{\mathcal{I}}}
\def\cJ{{\mathcal{J}}}

\def\eqref#1{{(\ref{#1})}}

\title{Chiral Symmetry Restoration from a Boundary}
\author{B.~C.~Tiburzi}
\email[]{btiburzi@ccny.cuny.edu}
\affiliation{Department of Physics,
        The City College of New York,  
        New York, NY 10031, USA}
\affiliation{
Graduate School and University Center,
        The City University of New York,
        New York, NY 10016, USA}
\affiliation{
RIKEN BNL Research Center, 
        Brookhaven National Laboratory, 
        Upton, NY 11973, USA}
\affiliation{
Kavli Institute for Theoretical Physics, 
University of California, Santa Barbara, CA 93106, USA}
\date{\today}
\begin{abstract}
The boundary of a manifold can alter the phase of a theory in the bulk. 
We explore the possibility of a boundary-induced phase transition for the chiral symmetry of QCD. 
In particular, 
we investigate the consequences of imposing homogeneous Dirichlet boundary conditions on the quark fields. 
Such boundary conditions are sometimes employed in lattice gauge theory computations, 
for example, when including external electromagnetic fields, 
or when computing quark propagators with a reduced temporal extent. 
Homogeneous Dirichlet boundary conditions force the chiral condensate to vanish at the boundary, 
and thereby obstruct the spontaneous breaking of chiral symmetry in the bulk. 
We show the restoration of chiral symmetry due to a boundary is a non-perturbative phenomenon
depending upon the mechanism of spontaneous symmetry breaking, 
and utilize the sigma model to exemplify the issues. 
Within this model, 
we find that chiral symmetry is completely restored if the length of the compact direction is less than
$2.0 \, \texttt{fm}$. 
For lengths greater than about
$4 \, \texttt{fm}$, 
an approximately uniform chiral condensate forms centered about the midpoint of the compact direction. 
While the volume-averaged condensate approaches the infinite volume value as the compact direction becomes very long, 
the finite-size corrections are shown to be power law rather than exponential. 
\end{abstract}
\maketitle

%%%%%%%%%%%%
\section{Introduction} %
%%%%%%%%%%%%

One of the hallmark non-perturbative features of QCD is spontaneous breaking of chiral symmetry. 
While quark masses explicitly break the chiral symmetry of the action, 
the lightest quarks 
(up and down)
have masses that can be treated as a perturbation about the 
symmetric
$SU(2)_L \otimes SU(2)_R$
chiral limit. 
The formation of a chiral condensate by the QCD vacuum in the chiral limit, 
namely $< \ol \psi \psi > \neq 0$, 
spontaneously breaks the chiral symmetry down to the vector subgroup, 
$SU(2)_V$. 
This symmetry breaking pattern along with the explicit breaking due to the quark masses gives an explanation of the lightness of the isotriplet of pseudoscalar pions because they must be the emergent Goldstone bosons.

Lattice gauge theory provides a first principles method for solving QCD numerically on finite Euclidean space-time lattices.
Strictly speaking,  
spontaneous symmetry breaking cannot occur in a finite volume%
~\cite{Weinberg:1996kr}. 
In practice, 
the formation of a chiral condensate on periodic lattices is determined by the size of the pion Compton wavelength compared to the lattice size%
~\cite{Gasser:1987zq}. 
When the pion Compton wavelength is small compared to the lattice size, 
$1/m_\pi \ll L$, 
the finite volume effect on the condensate is exponentially small and can be computed in chiral perturbation theory%
~\cite{Gasser:1986vb}. 
On the other hand, 
when the pion Compton wavelength is large compared to the lattice size, 
$1/m_\pi \gg L$, 
the zero-momentum mode of the coset of Goldstone fields becomes strongly coupled%
~\cite{Gasser:1987ah}.
Taking the chiral limit at finite volume puts one in the latter regime, 
which is the regime in which chiral symmetry is restored.

In this work, 
we explore a different restoration of chiral symmetry.%
\footnote{
Throughout we refer exclusively to the non-singlet chiral symmetry, 
$SU(2)_L \otimes SU(2)_R$. 
For the singlet case, 
it is convenient to phrase the discussion in terms of the axial symmetry of the QCD action, 
$U(1)_A$, 
which is anomalous in infinite volume. 
On a manifold with a boundary, 
the pressence of the axial anomaly is subtle%
~\cite{Atiyah:1975jf}.
With a homogeneous Dirichlet boundary, 
it is known that the integral of the singlet axial current's divergence vanishes% 
~\cite{Ninomiya:1984ge}. 
We leave the investigation of the chiral anomaly to future work. 
} 
We consider the fate of chiral symmetry on a Euclidean manifold with three infinite directions, 
and one compact direction that, unlike the periodic case, has a boundary. 
Specifically the compact direction is subject to homogeneous Dirichlet boundary conditions.  
Such boundary conditions are sometimes employed in lattice gauge theory computations.%
\footnote{
Inhomogeneous Dirichlet boundary conditions are imposed in the Schr\"odinger functional representation of QCD%
~\cite{Sint:1993un}. 
These boundary conditions too can affect spontaneous chiral symmetry breaking as will be made clear in a footnote below. 
} 
One example occurs in the study of hadron properties in external electromagnetic fields. 
Na{\"i}ve implementation of uniform external fields via linearly rising four-vector potentials leads to field gradients at 
the lattice boundary, 
and homogeneous Dirichlet boundary conditions have been sought to mitigate the effect of these electromagnetic field gradients on the quarks~\cite{Fiebig:1988en,Lee:2005ds,Christensen:2004ca,Lee:2005dq}. 
Additionally temporal Dirichlet boundary conditions have been imposed on lattices to compute quark propagators with a reduced temporal extent 
(so-called chopping of lattices)~\cite{Edwards:2005ym,Engelhardt:2007ub,Hagler:2007xi}.%
\footnote{
In both of these scenarios, 
only the valence quarks are subjected to Dirichlet boundary conditions, 
and one should formulate the problem in terms of graded symmetries~\cite{Sharpe:2000bc,Sharpe:2001fh}. 
For a partially quenched theory utilizing the same boundary conditions in the valence and sea sectors, 
chiral symmetry breaking takes the form
$SU(4|2)_L \otimes SU(4|2)_R \to SU(4|2)_V$. 
With different boundary conditions, 
however, 
there is no symmetry relating the valence and sea sectors,
and the chiral symmetry of the action must have the form 
$SU(2|2)_L \otimes SU(2)_L \otimes SU(2|2)_R \otimes SU(2)_R$, 
which is the same symmetry as in mixed-action lattice QCD calculations~\cite{Bar:2002nr}.
This reduced symmetry can break either to 
$SU(2|2)_V \otimes SU(2)_V$,
by the formation of valence and sea quark chiral condensates, 
or to
$SU(2|2)_L \otimes SU(2|2)_R \otimes SU(2)_V$, 
by the formation of only a sea quark chiral condensate. 
Our consideration above is for the valence subgroup of the graded chiral symmetry, 
i.e.~$SU(2)_L \otimes SU(2)_R  \subset SU(2|2)_L \otimes SU(2|2)_R$, 
and the mesons of the sigma model should be viewed as built only from valence quarks. 
One can go further and generalize the sigma model to appropriately reflect the graded symmetries;
but, 
this is beyond the scope of our work.  
}

The possibility that the boundary of a manifold can affect the phase of a theory in the bulk is a known phenomenon in condensed matter physics, 
for an overview of boundary critical phenomena, 
see% 
~\cite{Diehl:1996kd}. 
There has been some very recent work using the Gross--Neveu model to investigate the relation between chiral symmetry breaking and the Casimir force through a direct computation of the boundary separation and temperature dependence of the model's vacuum energy%
~\cite{Flachi:2012pf,Flachi:2013bc}.
To our knowledge, 
the effect of homogeneous Dirichlet boundary conditions on the chiral condensate in QCD has otherwise not been explored.

We employ a different approach to demonstrate that homogeneous Dirichlet boundary conditions can lead to the restoration of chiral symmetry.
Our consideration is in the absence of any external electromagnetic fields.    
Unlike on periodic lattices, 
we find that the restoration of chiral symmetry in the case of Dirichlet boundary conditions is not controlled by the size of the pion Compton wavelength. 
Instead,
the restoration of chiral symmetry in this case depends on the underlying mechanism of spontaneous symmetry breaking. 
To investigate these effects, 
we utilize the sigma model and constrain the model parameters from phenomenology, 
using some assumptions about the quark content of the lowest resonance in QCD. 
While there is rich phenomenology exploring the role isoscalar scalar states play in the mechanism of spontaneous chiral symmetry breaking, 
see~\cite{Fariborz:2008bd,Fariborz:2009cq}
and references therein, 
we employ the simplest possible model. 
In the sigma model, 
the Compton wavelength of the sigma meson strongly controls the restoration of chiral symmetry in the presence of a Dirichlet boundary. 
We find substantial effects on the chiral condensate for lattice sizes less than about
$4 \, \texttt{fm}$. 
As our consideration lies outside of chiral perturbation theory,  
our estimate is necessarily model dependent. 
The sigma model employed, 
however,  
has the minimal features necessary to address the effect, 
whereas chiral perturbation theory by contrast does not.  
An important finding within this model is that finite-size effects on the volume-averaged condensate only vanish as a power law rather than as an exponential.

Our presentation has the following organization. 
We first describe the simple sigma model that we employ in Sec.~\ref{s:Model}.  
The model is then considered in Sec.~\ref{s:D} with homogeneous Dirichlet boundary conditions in one of the space-time directions.  
We obtain an expression for the local chiral condensate by solving the analytical mechanics that determines the vacuum. 
The volume-averaged condensate is shown to approach the infinite volume value up to power-law corrections. 
Finally in Sec.~\ref{s:S}, 
a brief summary concludes this work.

%%%%%%%%%%%%%%%%%%%
\section{Sigma Model} \label{s:Model}%
%%%%%%%%%%%%%%%%%%

We begin by describing the simple sigma model we employ.
The ingredients are chosen so that the model shares the same pattern of symmetry breaking as QCD with two light flavors, 
and contains the simplest possible mechanism of spontaneous symmetry breaking.
Our consideration in this section is restricted to infinite volume.

The sigma model consists of an isoscalar scalar field, 
$S$, 
and an isovector pseudoscalar field,
$\vec{P}$.
These fields appear in the Euclidean space action density having the form
\begin{eqnarray}
\cL_E
&=& 
\frac{1}{2} \partial_\mu S \, \partial_\mu S
+ 
\frac{1}{2} \partial_\mu \vec{P} \cdot \partial_\mu \vec{P}
- 
\frac{\lambda m}{v}  S
\notag
\\
&&+
\Lambda 
\left( 
S^2 + \vec{P} \, {}^2 - v^2 
\right)^2  
\label{eq:sigma}
.\end{eqnarray}
There are four parameters of this model, 
$m$, 
$v$,
$\lambda$, 
and
$\Lambda$. 
The parameter 
$m$
is analogous to the quark mass. 
When the quark mass parameter is set to zero, 
the action density in 
Eq.~\eqref{eq:sigma} has an 
$SO(4) \cong SU(2) \otimes SU(2)$ 
symmetry. 
The quartic interaction term, 
however, 
leads to spontaneous breaking of the 
$SO(4)$  
symmetry down to 
$SO(3) \cong SU(2)$, 
by the formation of a condensate,
namely
$S_0^2 + \vec{P}_0^{\, 2} = v^2$.  
We append a subscript 
$``0"$ 
throughout to denote the vacuum expectation value of a field. 
Here the vacuum expectation values are uniform in space-time, 
e.g.~$S_0(x) = S_0$.

In discussing the sigma model, 
it is convenient to use a polar decomposition of the fields. 
To this end, 
we write
\begin{equation}
S + i \vec{P} \cdot \vec{\tau}= \Sigma U,
\end{equation}
where 
$\vec{\tau}$
are the usual isospin matrices, 
$\Sigma$
is a real-valued field, 
and
$U$ is a unitary field. 
The latter encompasses the Goldstone bosons
$\vec{\pi}$
in the form
$U = \exp ( i \vec{\pi} \cdot \vec{\tau} / F )$. 
In terms of the polar decomposition, 
the sigma model action density has the form
\begin{eqnarray}
\cL_E
&=& 
\frac{1}{4} 
\tr 
\left[ 
\partial_\mu \Sigma \partial_\mu \Sigma 
+ 
\Sigma^2 \partial_\mu U \partial_\mu U^\dagger 
\right]
\notag \\
&&- 
\frac{\lambda}{4 v}
\tr 
\left[
m \Sigma 
\left(
U + U^\dagger
\right)
\right]
+
\Lambda 
(\Sigma^2 - v^2)^2
.\end{eqnarray}
Traces appearing above are taken over isospin. 
The vacuum state of the theory is determined by minimizing the action density. 
The action has a minimum for the values:%
\footnote{
In the chiral limit, 
one has two possibilities for the vacuum expectation value of the sigma field, 
$\Sigma_0 = \pm v$. 
Inclusion of a small quark mass term, 
and the subsequent chiral limit, 
$m \to 0$,
will lead to the selection of 
$\Sigma_0 = + v$. 
} 
$U_0 = 1$
and
$\Sigma_0 = v$. Expanding about these vacuum expectation values, 
$U = 1 + \frac{i \vec{\pi} \cdot \vec{\tau}}{F} + \cdots$, 
and
$\Sigma = v + \sigma$, 
we have to quadratic order in the fields
\begin{equation}
\cL
=
\frac{1}{2} \partial_\mu \sigma \partial_\mu \sigma + \frac{1}{2} m_\sigma^2 \sigma^2
+
\frac{1}{2} \partial_\mu \vec{\pi} \cdot \partial_\mu \vec{\pi}
+ 
\frac{1}{2} m_\pi^2 \vec{\pi} \cdot \vec{\pi}
,\end{equation}
with the meson masses identified as
$m_\sigma^2 = 8 \Lambda F^2$,
and
$m_\pi^2 = \lambda m / F^2$.
Notice we must have the equality
$F = v$
in order for the pion kinetic term to have the proper canonical normalization.

The chiral condensate, 
which we denote by 
$< \ol \psi \psi >$, 
can be found by differentiating the vacuum energy density with respect to the explicit symmetry breaking parameter, 
which is the quark mass
$m$. 
The value of the chiral condensate is simply
\begin{equation}
< \ol \psi \psi >= - \lambda
.\end{equation}
This identification enables us to rewrite the pion mass in the form 
$F^2 m_\pi^2 = |< \ol \psi \psi >| \, m$, 
which is the Gell-Mann--Oakes--Renner relation. 
In the absence of explicit symmetry breaking, 
i.e.~the chiral limit when
$m = 0$, 
the pions are massless Goldstone modes. 
The sigma meson, 
on the other hand, 
remains massive, 
with its mass scale set by the pion decay constant 
$F$. 
In the chiral limit, 
the sigma meson can thus be integrated out to produce a low-energy theory of just the Goldstone modes%
~\cite{Ecker:1988te}. 
This theory is chiral perturbation theory and is a model independent description of low-energy QCD%
~\cite{Gasser:1983yg}. 
To study the effects of a Dirichlet boundary on the chiral condensate, 
however, 
we must retain the sigma degree of freedom. 
Without the sigma, 
the chiral limit condensate would be frozen into a uniform value, 
and this conflicts with the quark boundary conditions.

Using some assumptions,  
we can use phenomenology to fix the model parameters. 
With our normalization, 
the vacuum expectation value
$v$ 
of the sigma field is identical to the parameter
$F$, 
which can be identified as the pion decay constant once an external axial-vector field has been coupled to the theory. 
Technically 
$F$ 
is the chiral limit value of the pion decay constant, 
and has the approximate numerical value
$F = 93 \, \texttt{MeV}$. 
We assume that the sigma field corresponds to the lightest resonance in QCD. 
The mass of this sigma resonance has been determined from a detailed analysis of high-precision $\pi \pi$ scattering data%
~\cite{Caprini:2005zr}.  
From the extracted mass of the sigma, 
$m_\sigma = 440 \, \texttt{MeV}$, 
we find the parameter
$\Lambda$
has the value
$\Lambda = 2.8$. 
In our model, 
the sigma meson is a particle not a resonance, 
and consequently the usual caveats about meson models apply. 
The chiral condensate can be used to fix the value of the parameter 
$\lambda$, 
which itself is the chiral limit value of the condensate up to sign. 
This quantity, 
however, 
is QCD renormalization scale and scheme dependent, 
and we simply choose to quote results in terms of ratios to the infinite volume condensate,  
for which the scale and scheme dependence exactly cancels.%
\footnote{
It is efficacious to spell out a procedure to investigate the effects of Dirichlet boundary conditions on the chiral condensate using the lattice as a regulator. 
In this case, 
the condensate is both multiplicatively and additively renormalized at finite values of the quark mass.
We can define a subtracted chiral condensate for the case of Dirichlet boundary conditions in a way that is analogous 
to the temperature dependent case, 
see, for example,%
~\cite{Petreczky:2012rq}. 
Consider two values of the quark mass, 
$m_1$, 
and 
$m_2$
for a given lattice action. 
Using this ultraviolet regulator, 
one computes four quantities: 
the chiral condensate with periodic boundary conditions for each quark mass, 
$< \ol \psi \psi >_{m_1}$ 
and
$< \ol \psi \psi >_{m_2}$, 
and the local chiral condensate with Dirichlet boundary conditions for each quark mass, 
$< \ol \psi \psi (x) >_{m_1}$ 
and
$< \ol \psi \psi (x)>_{m_2}$. 
The subtracted condensates, 
$< \ol \psi \psi >_{\text{sub}} = < \ol \psi \psi >_{m_1} - \frac{m_1}{m_2} < \ol \psi \psi >_{m_2}$
and
$< \ol \psi \psi(x) >_{\text{sub}} = < \ol \psi \psi (x)>_{m_1} - \frac{m_1}{m_2} < \ol \psi \psi (x) >_{m_2}$, 
are free of power-law divergences, 
which are proportional to
$m \, a^{-2}$. 
In the latter case, 
the subtraction cancels a quark-mass independent function that depends on 
$x$.
The ratio of subtracted condensates, 
$< \ol \psi \psi(x) >_{\text{sub}} / < \ol \psi \psi >_{\text{sub}}$, 
can then be used to study the behavior of the chiral condensate in the infrared. 
Because the chiral condensate calculated with periodic boundary conditions suffers from finite-size effects, 
one should first take the infinite volume limit of     
$< \ol \psi \psi >_{\text{sub}}$
in order to isolate the finite-size effects arising solely from Dirichlet boundary conditions when forming the ratio. 
Finally
the local chiral condensates can be averaged over the compact direction to investigate the behavior of the volume-averaged condensate. 
Such quantities are more likely to be computed using the lattice in order to reduce statistical noise. 
}

%%%%%%%%%%%%%%%%%%%%%
\section{Dirichlet Boundary} \label{s:D}%
%%%%%%%%%%%%%%%%$%%%

Above we describe the symmetry breaking in the sigma model in infinite volume. 
To minimize the action as a functional of the meson fields, 
we eliminated the kinetic terms by restricting our attention to uniform field configurations. 
These uniform field configurations are energetically preferred because the kinetic energy contributes
positively to the action density. 
With a homogeneous Dirichlet boundary condition, 
however, 
there are no non-trivial field configurations that are uniform. 
On the other hand, 
the potential energy terms in the action drive the 
$\Sigma$
field to a non-vanishing vacuum expectation value. 
In general, 
the Dirichlet boundary now allows for competition between the kinetic and potential terms, 
and the symmetry breaking pattern of the model must be scrutinized.

For simplicity, 
we choose to impose a homogeneous Dirichlet boundary condition in just one of the space-time directions. 
Because we work in Euclidean space, 
we need not specify whether it is a spatial or temporal direction that has a boundary. 
The remaining three directions are kept infinite.
We thus consider the quark field 
$\psi(x)$
to satisfy the boundary conditions
$\psi(x =0) = 0$, 
and
$\psi (x = L) = 0$, 
where 
$L$ 
is the length of the $x$-direction.%
\footnote{
In the Sch\"odinger functional representation of QCD%
~\cite{Sint:1993un}, 
one imposes inhomogeneous Dirichlet boundary conditions in the temporal direction having the form,
%\begin{equation}
%\notag
$\cP_+ \psi(x_4=0) = \ol \psi (x_4=0) \cP_- = 0$, 
%\end{equation}
and
%\begin{equation}
%\notag
$\cP_- \psi (x_4= T) = \ol \psi(x_4=T) \cP_+ = 0$,
%\end{equation} 
where
$\cP_\pm = \frac{1}{2} ( 1 \pm \gamma_4)$
are parity projection matrices, 
and the remaining components of the fermion field are non-vanishing at the boundaries. 
From a renormalization group argument, 
such boundary conditions are natural in the continuum limit%
~\cite{Luscher:2006df}.
The chiral condensate can be written trivially as the sum,
%\begin{equation}
%\notag
$< \ol \psi \psi (x)> 
= 
< \ol \psi \cP_+ \psi (x) > 
+ 
< \ol \psi \cP_- \psi (x) >$,
%\end{equation} 
which consequently vanishes at the boundaries
$x_4 = 0$, 
and 
$T$. 
The Schr\"odinger functional is largely employed as a renormalization scheme; 
and, 
as a massless scheme, 
the massless fermions cannot scatter off the chiral condensate. 
Restoration of chiral symmetry due to the boundary is then irrelevant for determining the renormalization of operators in this scheme. 
The same is not true for the calculation of hadron properties at non-vanishing quark masses within the Schr\"odinger functional formulation. 
While the utilization of small temporal lattices becomes possible to compute hadron properties, 
see, for example, 
\cite{Guagnelli:1999zf}, 
one must be careful to account for possible non-perturbative finite-size effects resulting from the temporal dependence of the chiral condensate. 
}
The other space-time directions are treated implicitly in our notation.

The imposition of quark boundary conditions translates into boundary conditions on the 
$\S$ 
and 
$\vec{\pi}$
fields of the sigma model.
We assume these meson fields have the same quantum numbers of certain quark-level interpolating operators.%
\footnote{
While such an interpolating operator for the 
$\Sigma$ 
field is perfectly reasonable in the context of this model, 
it must be mentioned that the isoscalar scalar resonances in QCD exhibit an unusual spectroscopy.
In particular the inverted level ordering of the states is suggestive of a two-quark, two-antiquark structure, 
as first pointed out in~\cite{Jaffe:1976ig}. 
An admixture of such states in the physical sigma resonance should not pose a problem for our results
because local two-quark, two-antiquark interpolating operators
also vanish at a Dirichlet boundary. 
Mixing with a scalar glueball, 
however, would complicate things. 
For quark-antiquark scalars, 
such mixing is not possible in the chiral limit, 
while mixing becomes possible for certain two-quark, two-antiquark 
scalars.  
If our field 
$\Sigma$
were to contain a small scalar glueball component; 
then, 
by virtue of the gauge field periodicity,
the chiral condensate would not vanish at the boundary, 
see Eq.~\eqref{eq:sigcond}, 
which would contradict the quark boundary conditions. 
In this case, 
the glueball component of the
$\Sigma$ 
field would need to be removed in order to be consistent. 
Such issues are important to incorporate in a more realistic model of spontaneous chiral symmetry breaking. 
} 
In particular, 
we take
$\S(x) \sim \ol \psi(x) \psi(x)$
and 
$\vec{\pi}(x) \sim \ol \psi(x) \gamma_5  \vec{\tau} \, \psi(x)$. 
As a result of these identifications,
we have the boundary conditions
\begin{eqnarray}
\Sigma(x = 0) &=& \Sigma( x = L) = 0,
\notag \\
U(x=0) &=& U(x=L) = 1
\label{eq:DBCs}
.\end{eqnarray}
These boundary conditions can be directly obtained if one derives the sigma model from a model with quarks, 
namely from the Nambu--Jona-Lasinio model~\cite{Ebert:1985kz}.

To find the vacuum of the theory with Dirichlet boundary, 
we minimize the action density. 
For the directions of infinite extent, 
the fields are frozen into their corresponding zero momentum modes. 
On account of Eq.~\eqref{eq:DBCs}, 
the unitary field 
$U$
can take on a uniform value
$U_0 = 1$
for all space-time. 
The quark mass term will consequently be minimized as a function of 
$U_0$
for 
$U_0 = 1$
provided the vacuum value of the field 
$\Sigma$ 
is positive. 
As the quark mass dependent case is more complicated, 
we will work in the chiral limit, 
$m=0$, 
for which the vacuum minimization requires 
$U_0 =1$. 
It is important to note that adding a small quark mass term to the action leads to an energetic preference for positive values of
$\Sigma$.

The vacuum value of the field 
$\Sigma$
depends on the $x$-coordinate, 
namely 
$\Sigma = \Sigma_0(x) + \cdots$. 
In the chiral limit, 
this vacuum expectation value will minimize the action density
\begin{equation}
S[\Sigma_0] 
= 
\frac{1}{L}
\int_0^L 
\Bigg[
\frac{1}{2} 
\left( 
\frac{d \Sigma_0}{dx} 
\right)^2
+
\Lambda
\left( 
\Sigma_0^2 
- 
v^2 
\right)^2
\Bigg]
dx
,\end{equation}
which is a functional of 
$\Sigma_0(x)$. 
The functional minimization, 
%\begin{equation}
$\delta S /\delta \Sigma_0 (x)=0$,
%\end{equation}
can be achieved through solving the Euler-Lagrange equation. 
Once this solution is known, 
the value of the chiral condensate follows immediately
\begin{equation} \label{eq:sigcond}
< \ol \psi \psi (x) > 
= - \frac{\lambda}{F} \Sigma_0 (x) 
.\end{equation}
Accordingly the chiral condensate will vanish at the boundary, 
but could develop a non-zero value in the bulk. 
It remains to solve the Euler-Lagrange equation to determine 
$\Sigma_0(x)$.

The problem of minimizing the action density is equivalent to a problem in analytical mechanics. 
The mechanical analogue of energy is given by
\begin{equation}
\mathcal{E}
=
\frac{1}{2} 
\left( 
\frac{d \Sigma_0}{dx} 
\right)^2
- 
\Lambda
\left( 
\Sigma_0^2 - v^2
\right)^2
,\end{equation}
and is a constant of the motion. 
Because of the Dirichlet boundary at 
$x = 0$ 
and 
$x = L$, 
any non-trivial solution for 
$\Sigma_0(x)$
must have at least one turning point. 
The turning points 
$x_j$
are determined by the condition
$d \Sigma_0 / d x \big|_{x = x_j} = 0$. 
Labeling the corresponding values of the field at the turning points, 
$\Sigma_0^{(j)} = \Sigma_0 ( x_j)$, 
we see that the latter are given by
\begin{eqnarray}
\Sigma_0^{(1)} 
&=&
+v
\sqrt{1 - \xi}
,
\qquad
\Sigma_0^{(2)} 
=
-v
\sqrt{1 - \xi} 
,\notag \\
\Sigma_0^{(3)} 
&=&
-v
\sqrt{1+ \xi}
,\qquad 
\Sigma_0^{(4)} 
=
+v
\sqrt{1 + \xi} 
,\end{eqnarray}
with 
$\xi = \sqrt{ - \frac{\mathcal{E}}{\Lambda v^2}}$. 
In order for there to be any turning points at all, 
it must be the case that 
$\mathcal{E} < 0$. 
For the first two turning points to exist, 
one additionally requires
$\xi < 1$.

The concavity of the function 
$\Sigma_0(x)$
follows from the analogue of Newton's second law. 
The force is given by 
$F = -d V/d \Sigma_0$,
with 
$V = - \Lambda ( \Sigma_0^2 - v^2)^2$, 
and implies that the values of the field at the turning points,
$\Sigma_0^{(j)}$,
are local minima for 
$j$ 
even,  
and local maxima for 
$j$
odd.  
As a result, 
we need not consider the turning points 
$\Sigma_0^{(3)}$
and 
$\Sigma_0^{(4)}$
in determining the solution 
$\Sigma_0(x)$.  
For example, 
consider a solution which rises from
$\Sigma_0 = 0$ 
at 
$x = 0$. 
For this solution to turn over, 
we need a positive turning point corresponding to a maximum. 
The only possibility is the value
$\Sigma_0^{(1)}$. 
The solution
$\Sigma_0(x)$ 
could then decrease from the maximum down to a minimum before rising again, 
but the only possible turning point that corresponds to a minimum with value less than 
$\Sigma_0^{(1)}$ 
is 
$\Sigma_0^{(2)}$. 
The solution 
$\Sigma_0(x)$
is necessarily bounded by 
$\Sigma_0^{(1)}$ 
from above and
$\Sigma_0^{(2)}$ 
from below. 
There are an infinite number of solutions which minimize the action. 
They are characterized by the number of oscillations between the extrema
$\Sigma_0^{(1)}$ 
and
$\Sigma_0^{(2)}$. 
Because the latter is negative,  
the inclusion of a small quark mass term will lead to an energetic disadvantage for all solutions having any turning points for which 
$\Sigma_0(x)$ 
attains the value
$\Sigma_0^{(2)}$.

%%%%%%%%%%%%%%%%%%%%%%%%%%%%%%%%%%%%%%
%
%%%%%%%%%%%%%%%%%%%%%%%%%%%%%%%%%%%%%%
%
%%%%%%%%%%%%%%%%%%%%%%%%%%%%%%%%%%%%%%
\begin{figure}[t]
\epsfig{file=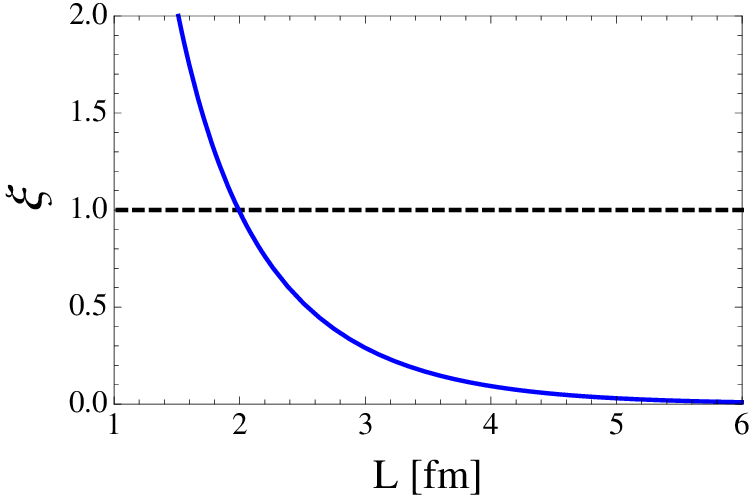,width=0.4\textwidth}
\caption{
The solution 
$\xi$ 
as a function of the finite extent 
$L$ 
of the 
$x$-direction. 
Values of 
$L$ 
for which
$\xi \geq 1$
lead to the vacuum expectation value
$\Sigma_0(x) = 0$, 
and hence correspond to a complete restoration of chiral symmetry in the sigma model.
}
\label{f:xi}
\end{figure}
%%%%%%%%%%%%%%%%%%%%%%%%%%%%%%%%%%%%%
%
%%%%%%%%%%%%%%%%%%%%%%%%%%%%%%%%%%%%%
%
%%%%%%%%%%%%%%%%%%%%%%%%%%%%%%%%%%%%%

The solution 
$\Sigma_0(x)$
we seek can thus be characterized as monotonically increasing from zero up to the maximum value
$\Sigma_0^{(1)}$,  
and then monotonically decreasing back down to zero. 
The motion of 
$\Sigma_0$ 
is symmetric about the turning point; 
consequently, 
we must have
$x_1 = L / 2$.  
Integrating the equation of motion from the boundary to the turning point (or vice versa), 
we arrive at the equation
\begin{eqnarray} \label{eq:trans}
\texttt{K} \left( 
\frac{1 - \xi}{1 + \xi}
\right)
=  
\frac{\ell}{2}
\sqrt{1 + \xi}
,\end{eqnarray}
where
$\texttt{K}(m)$ 
is the complete elliptic integral of the first kind. 
This is a special case of the incomplete elliptic integral of the first kind, 
$\texttt{F}(\phi \, | m)$, 
defined by
\begin{equation}
\texttt{F}(\phi \, | m) = \int_0^{\phi}  \frac{d\theta}{\sqrt{1 - m \sin^2 \theta}}
,\end{equation} 
namely
$\texttt{K}(m) = \texttt{F}(\frac{\pi}{2} | m)$.
Above, 
we have employed the abbreviation
\begin{equation}
\ell = v L \sqrt{2 \Lambda} = \frac{1}{2} m_\sigma L
.\end{equation}  
The relation expressed in Eq.~\eqref{eq:trans} 
implicitly defines the analogue of the mechanical energy 
$\mathcal{E}$
as a function of the sigma model parameters and the extent of the $x$-direction, 
i.e.~%
$\mathcal{E} = \mathcal{E} (v, \Lambda,  L)$. 
In practice, 
it is simpler to work with the dimensionless variable
$\xi = \xi ( v, \Lambda, L)$. 
In Fig.~\ref{f:xi}, 
we plot the value of 
$\xi$
that satisfies 
Eq.~\eqref{eq:trans}
as a function of the extent of the compact direction, 
$L$.
For 
\begin{equation}
L 
< 
\frac{\pi}{2 v \sqrt{\Lambda}}
=
\frac{\sqrt{2} \, \pi}{m_\sigma} 
\approx 2.0 \, \texttt{fm}
,\end{equation} 
the solution requires
$\xi \geq 1$
for which the turning point
$\Sigma_0^{(1)}$
does not exist, 
and consequently chiral symmetry is completely restored in the model.

With the value of 
$\xi$ 
determined from Eq.~\eqref{eq:trans}, 
we can implicitly specify the solution 
$\Sigma_0(x)$
by integrating the equation of motion to an arbitrary point 
$x$. 
We find the solution must satisfy
\begin{multline} \label{eq:IT}
\texttt{F}
\left( 
\sin^{-1} \frac{\Sigma_0}{\Sigma_0^{(1)}}
\Bigg| 
\frac{1 - \xi}{1 + \xi}
\right)
= 
\ell \sqrt{1 + \xi}
\\
\times
\begin{cases}
\frac{x}{L}, 
& \text{for} 
\quad 0 \leq x \leq \frac{L}{2}
\\
1 - \frac{x}{L}, 
& \text{for} 
\quad \frac{L}{2} \leq x \leq L
\end{cases}
,\end{multline}
with 
$\texttt{F}(\phi \, | m)$
the incomplete elliptic integral defined above. 
Notice the full solution agrees with Eq.~\eqref{eq:trans} at the turning point.

%%%%%%%%%%%%%%%%%%%%%%%%%%%%%%%%%%%%%%
%
%%%%%%%%%%%%%%%%%%%%%%%%%%%%%%%%%%%%%%
%
%%%%%%%%%%%%%%%%%%%%%%%%%%%%%%%%%%%%%%
\begin{figure}[b]
\epsfig{file=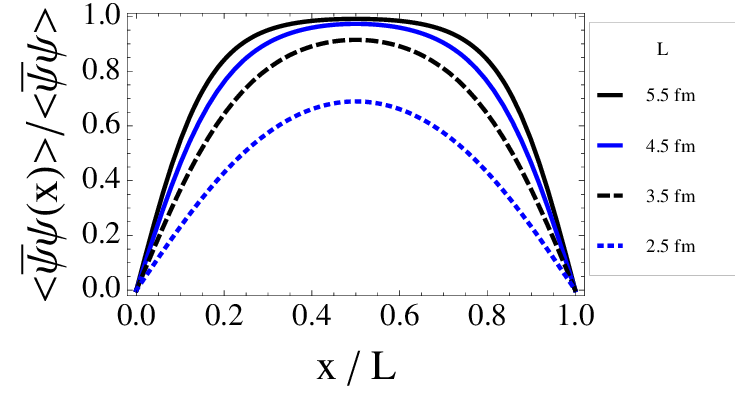,width=0.48\textwidth}
\caption{
Ratio of the chiral condensate with homogeneous Dirichlet boundary conditions to the infinite volume chiral condensate 
plotted as a function of the $x$-coordinate scaled by 
$L$.
For 
$L < 2.0 \, \texttt{fm}$,  
the condensate vanishes everywhere, 
$< \ol \psi \psi (x) > = 0$. 
}
\label{f:chi}
\end{figure}
%%%%%%%%%%%%%%%%%%%%%%%%%%%%%%%%%%%%%
%
%%%%%%%%%%%%%%%%%%%%%%%%%%%%%%%%%%%%%
%
%%%%%%%%%%%%%%%%%%%%%%%%%%%%%%%%%%%%%

From this solution, 
we can determine the chiral condensate as a function of the $x$-coordinate. 
It is simplest to consider the ratio of the condensate with Dirichlet boundary conditions with that of the condensate in infinite volume, 
namely
$< \ol \psi \psi (x) > / < \ol \psi \psi >$. 
This ratio is plotted in 
Fig.~\ref{f:chi}
for several values of the extent 
$L$. 
For 
$L < 2.0 \, \texttt{fm}$, 
the chiral condensate identically vanishes for all 
$x$. 
Above this size, 
a non-vanishing condensate forms, 
however, 
the condensate is significantly altered from its infinite volume value for sizes less that about 
$4 \, \texttt{fm}$. 
When 
$L = 3.5 \, \texttt{fm}$, 
for example, 
the condensate is reduced by more than 
$25 \%$
over half of the length of the $x$-direction. 
The condensate, 
moreover, 
reaches a maximum value that is 
$15 \%$ 
smaller than the infinite volume value. 
For sizes greater than about 
$4 \, \texttt{fm}$, 
an approximately uniform chiral condensate is formed. 
The support of the condensate, 
however,
is over roughly 
$50 \%$ 
of the extent of the $x$-direction, 
and is naturally centered about the midpoint 
$x = L /2$.

While 
the chiral condensate must vanish at the boundary for all 
$L$, 
chiral symmetry is spontaneously broken in the bulk of the lattice
in the limit of large 
$L$. 
 The maximum value of the chiral condensate approaches the infinite volume value,
because the sigma field at the turning point has the behavior,
$\Sigma_0^{(1)} \to v$
as 
$L$
goes to infinity. 
This behavior arises due to 
$\xi$
approaching zero, 
see Fig.~\ref{f:xi}. 
From Eq.~\eqref{eq:trans}, 
we can analytically derive the large 
$L$ 
limit of the maximum value of the condensate. 
Using the expansion of the complete elliptic integral about unity, 
we find the midpoint value is given by
\begin{equation}
< \ol \psi \psi (L/2) >
=
< \ol \psi \psi > 
\left[1 - 4 \exp \left( - \frac{ m_\sigma L}{2} \right) 
+ 
\cdots
\right]
,\end{equation}
which shows that the asymptotic behavior is controlled by the Compton wavelength of the sigma meson.

\begin{figure}
\epsfig{file=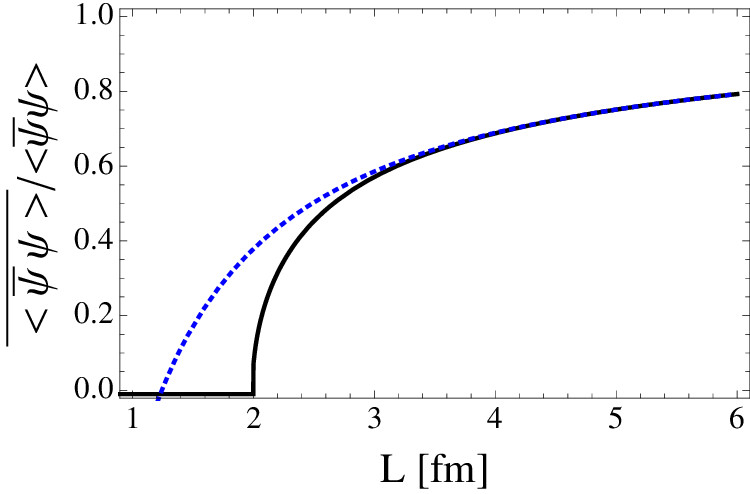,width=0.4\textwidth}
\caption{
Ratio of the volume-averaged condensate in the sigma model
$\overline{<\ol \psi \psi>}$ 
to the infinite volume condensate
$< \ol \psi \psi >$
plotted as a function of the finite extent 
$L$. 
The dotted curve shows the asymptotic formula given in 
Eq.~\eqref{eq:asym}.
}
\label{f:power}
\end{figure}

The value of the chiral condensate at the midpoint does not completely characterize the fate of chiral symmetry because there is also the issue of support over the bulk of the lattice. 
In addressing the extent to which chiral symmetry is restored, 
one can also study the chiral condensate averaged over the compact direction. 
This average is simply defined by
\begin{equation}
\overline{ < \ol \psi \psi > }
=
\frac{1}{L}
\int_0^L dx 
< \ol \psi \psi(x) > 
.\end{equation}
In the limit of an asymptotically large extent 
$L$, 
the volume-averaged chiral condensate tends to the infinite volume value, 
however, 
the approach to asymptopia is slow.
In the asymptotic limit, 
we expand the second argument of the incomplete elliptic integral in Eq.~\eqref{eq:IT} about unity. 
The asymptotic condensate can be determined in closed form, 
and averaged over the 
$x$-direction to produce
\begin{equation} \label{eq:asym}
\overline{ < \ol \psi \psi > }
= 
< \ol \psi \psi > 
\left( 
1
- 
\frac{4 \log 2}{m_\sigma L}
+ 
\cdots
\right)
.\end{equation} 
This power-law behavior is confirmed in Fig.~\ref{f:power}, 
where we compare the volume-averaged condensate determined from the full solution, 
Eq.~\eqref{eq:IT}, 
to the asymptotic form given in 
Eq.~\eqref{eq:asym}. 
While the asymptotic form works very well for 
$L > 3 \, \texttt{fm}$, 
the volume-averaged condensate only slowly approaches the infinite volume value.

%%%%%%%%%%%%%%%%%%%%%%%%%%%%%%%%%%%%%%
%
%%%%%%%%%%%%%%%%%%%%%%%%%%%%%%%%%%%%%%
%
%%%%%%%%%%%%%%%%%%%%%%%%%%%%%%%%%%%%%%
%%%%%%%%%%%%%%%%%%%%%%%%%%%%%%%%%%%%%
%
%%%%%%%%%%%%%%%%%%%%%%%%%%%%%%%%%%%%%
%
%%%%%%%%%%%%%%%%%%%%%%%%%%%%%%%%%%%%%

As a final remark in our discussion, 
we note that the volume-averaged condensate exhibits a cusp at the point where chiral symmetry breaking can occur. 
This sharp behavior is likely an artifact of the simplicity of the sigma model we employ. 
Because chiral dynamics of pions is insufficient to describe chiral symmetry breaking in the presence of a Dirichlet boundary, 
we include the lowest-lying state with the requisite quantum numbers to address the problem. 
Due to confinement,
however, 
there are a tower of such states, 
and vacuum expectation values of these fields will become relevant at short distances. 
Inclusion of a higher-lying scalar state 
$\Sigma'$
in the model will presumably smooth out the cusp seen around
$L = 2 \,\texttt{fm}$, 
and introduce a milder cusp at a smaller length scale, 
$L' =\sqrt{2} \, \pi / m_{\sigma'}$. 
We leave the inclusion of such states required in more realistic models of chiral symmetry breaking to future work.

%%%%%%%%%%%%%%%%%%%%%
\section{Summary} \label{s:S}              %
%%%%%%%%%%%%%%%%$%%%

Above we investigate the effect of homogeneous Dirichlet boundary conditions on the chiral condensate. 
This effect cannot be ascertained within chiral perturbation theory, 
because the chiral condensate is determined by the expression
\begin{equation}
< \ol \psi \psi (x) >  = - \frac{\lambda}{4} < U(x) + U^\dagger(x) > + \cdots
.\end{equation} 
The right-hand side of this relation does not vanish at the boundary in contradiction with the quark boundary conditions satisfied on the left-hand side. 
A consistent treatment of the chiral condensate in the presence of a Dirichlet boundary necessitates
including the dynamics of isoscalar scalar mesons. 
For this reason, 
we employ the sigma model which shares the same symmetry breaking pattern as QCD with two light quark flavors, 
and provides the simplest model of spontaneous chiral symmetry breaking. 
Identifying the sigma field with the sigma resonance of QCD,
model parameters are chosen to reproduce phenomenology. 
We then subject the model to homogeneous Dirichlet boundary conditions in one of the space-time directions. 
The chiral condensate is determined by minimizing the action to find the vacuum configuration, 
and this is achieved in the sigma model using well-known functions.

Chiral symmetry is shown to be restored in the sigma model for sizes less than
$2.0 \, \texttt{fm}$. 
For sizes greater than about 
$4 \, \texttt{fm}$, 
an approximately uniform condensate forms centered about the midpoint of the compact direction
and having support over roughly half of the compact direction. 
The volume-averaged condensate suffers from finite-size effects that are power law rather than exponential. 
As our estimation is necessarily model dependent, 
it would be useful to compare the behavior of the chiral condensate in quark models with spontaneous chiral symmetry breaking, 
and realistic meson models with resonances. 
Indeed, 
it is well recognized that the sigma meson and other isoscalar scalar resonances play a crucial role in  the mechanism of spontaneous chiral symmetry breaking. 
To understand the behavior of the chiral condensate with decreasing size
$L$, 
a two-flavor model is insufficient, 
and higher-lying scalar resonances should be included. 
The unusual spectroscopy of such states suggests that 
a nonet of two-quark, two-antiquark states may be needed in addition to a nonet of quark-antiquark states, 
see, 
for example,%
~\cite{Fariborz:2008bd,Fariborz:2009cq}. 
Inclusion of such states should not be problematic in our framework, 
however, 
further mixing with scalar glueballs would require a more careful treatment.

Nonetheless, 
we believe one should be cautious in interpreting results from lattice computations employing homogeneous
Dirichlet boundary conditions on lattices less than about 
$4 \, \texttt{fm}$ 
in extent. 
As our estimate, 
moreover, 
comes from treating only one space-time direction as compact,
it would be interesting to consider the case of two directions subject to Dirichlet boundaries. 
Such a scenario is encountered in some lattice computations with external electromagnetic fields%
~\cite{Alexandru:2010dx,Freeman:2012cy}. 
This setup would additionally allow for an exploration of the Casimir effect.  
The three-dimensional case is also of interest to confront phenomenology of the bag model%
~\cite{Chodos:1974je,Chodos:1974pn}, 
and the proposal of in-hadron condensates%
~\cite{Brodsky:2008xm,Brodsky:2008be}. 
Finally to establish the credibility of lattice computations with homogeneous Dirichlet boundaries, 
one requires a lattice computation of the chiral condensate,
either locally or volume averaged,  
which will ultimately reveal the extent to which chiral symmetry is restored in the presence of a boundary. 
In turn, 
frustration of the chiral condensate via Dirichlet boundaries may enable us to learn more about the mechanism that underlies spontaneous chiral symmetry breaking.

%%%%%%%%%%%%%     
\begin{acknowledgments}
Work supported in part by a joint CCNY--RBRC fellowship, 
a PSC-CUNY award, 
a CUNY-JFRASE award, 
and by the U.S.~National Science Foundation, 
under Grant No.~PHY$12$-$05778$.
Completion of this work was carried out at the KITP, 
whose hospitality and partial support through the
U.S.~National Science Foundation, 
under Grant No.~PHY$11$-$25915$,
are gratefully acknowledged. 
Furthermore, 
we thank 
W.~Detmold 
for useful comments. 
\end{acknowledgments}
%%%%%%%%%%%%

\appendix

\bibliography{hb}

\end{document}